\input harvmac

\input amssym

\def\unit{\relax{\rm 1\kern-.26em I}}
\def\nada{\relax{\rm 0\kern-.30em l}}
\def\tilde{\widetilde}
\def\t{\tilde}

\def\pl{M_{{\rm pl}}}


\def\det{{\rm det}}

\noblackbox
\def\IL{\relax{\rm I\kern-.18em L}}
\def\IH{\relax{\rm I\kern-.18em H}}
\def\IR{\relax{\rm I\kern-.18em R}}
\def\IC{\relax\hbox{$\inbar\kern-.3em{\rm C}$}}
\def\IZ{\relax\ifmmode\mathchoice
{\hbox{\cmss Z\kern-.4em Z}}{\hbox{\cmss Z\kern-.4em Z}} {\lower.9pt\hbox{\cmsss Z\kern-.4em Z}}
{\lower1.2pt\hbox{\cmsss Z\kern-.4em Z}}\else{\cmss Z\kern-.4em Z}\fi}

\def\CO {{\cal O}}
\def\CZ {{\cal Z}}


\def\CO {{\cal O}}

\def\CZ {{\cal Z }}

\def\det{{\rm det}}
\def\Tr{{\rm Tr}}

\font\manual=manfnt \def\dbend{\lower3.5pt\hbox{\manual\char127}}

\def\IZ{\relax\ifmmode\mathchoice
{\hbox{\cmss Z\kern-.4em Z}}{\hbox{\cmss Z\kern-.4em Z}} {\lower.9pt\hbox{\cmsss Z\kern-.4em Z}}
{\lower1.2pt\hbox{\cmsss Z\kern-.4em Z}}\else{\cmss Z\kern-.4em Z}\fi}

\def\lfm#1{\medskip\noindent\item{#1}}

\def\bar{\overline}

\def\rt2{\sqrt{2}}
\def\irt2{{1\over\sqrt{2}}}

\def\t{\tilde}

\def\slashchar#1{\setbox0=\hbox{$#1$}           
   \dimen0=\wd0                                 
   \setbox1=\hbox{/} \dimen1=\wd1               
   \ifdim\dimen0>\dimen1                        
      \rlap{\hbox to \dimen0{\hfil/\hfil}}      
      #1                                        
   \else                                        
      \rlap{\hbox to \dimen1{\hfil$#1$\hfil}}   
      /                                         
   \fi}

\def\foursqr#1#2{{\vcenter{\vbox{
    \hrule height.#2pt
    \hbox{\vrule width.#2pt height#1pt \kern#1pt
    \vrule width.#2pt}
    \hrule height.#2pt
    \hrule height.#2pt
    \hbox{\vrule width.#2pt height#1pt \kern#1pt
    \vrule width.#2pt}
    \hrule height.#2pt
        \hrule height.#2pt
    \hbox{\vrule width.#2pt height#1pt \kern#1pt
    \vrule width.#2pt}
    \hrule height.#2pt
        \hrule height.#2pt
    \hbox{\vrule width.#2pt height#1pt \kern#1pt
    \vrule width.#2pt}
    \hrule height.#2pt}}}}
\def\psqr#1#2{{\vcenter{\vbox{\hrule height.#2pt
    \hbox{\vrule width.#2pt height#1pt \kern#1pt
    \vrule width.#2pt}
    \hrule height.#2pt \hrule height.#2pt
    \hbox{\vrule width.#2pt height#1pt \kern#1pt
    \vrule width.#2pt}
    \hrule height.#2pt}}}}
\def\sqr#1#2{{\vcenter{\vbox{\hrule height.#2pt
    \hbox{\vrule width.#2pt height#1pt \kern#1pt
    \vrule width.#2pt}
    \hrule height.#2pt}}}}
\def\square{\mathchoice\sqr65\sqr65\sqr{2.1}3\sqr{1.5}3}

\def\figin{\epsfcheck\figin}\def\figins{\epsfcheck\figins}
\def\epsfcheck{\ifx\epsfbox\UnDeFiNeD
\message{(NO epsf.tex, FIGURES WILL BE IGNORED)}
\gdef\figin##1{\vskip2in}\gdef\figins##1{\hskip.5in}
\else\message{(FIGURES WILL BE INCLUDED)}%
\gdef\figin##1{##1}\gdef\figins##1{##1}\fi}
\def\DefWarn#1{}
\def\figinsert{\goodbreak\midinsert}
\def\ifig#1#2#3{\DefWarn#1\xdef#1{fig.~\the\figno}
\writedef{#1\leftbracket fig.\noexpand~\the\figno}%
\figinsert\figin{\centerline{#3}}\medskip\centerline{\vbox{\baselineskip12pt \advance\hsize by
-1truein\noindent\footnotefont{\bf Fig.~\the\figno:\ } \it#2}}
\bigskip\endinsert\global\advance\figno by1}


\lref\WittenNF{
 E.~Witten,
 ``Dynamical Breaking Of Supersymmetry,''
 Nucl.\ Phys.\ B {\bf 188}, 513 (1981).
}

\lref\GiudiceBP{
 G.~F.~Giudice and R.~Rattazzi,
 ``Theories with gauge-mediated supersymmetry breaking,''
 Phys.\ Rept.\ {\bf 322}, 419 (1999)
 [arXiv:hep-ph/9801271].
}

\lref\PoppitzVD{
  E.~Poppitz and S.~P.~Trivedi,
  ``Dynamical supersymmetry breaking,''
  Ann.\ Rev.\ Nucl.\ Part.\ Sci.\  {\bf 48}, 307 (1998)
  [arXiv:hep-th/9803107].
}

\lref\TerningTH{
  J.~Terning,
  ``Non-perturbative supersymmetry,''
  arXiv:hep-th/0306119.
}

\lref\ShadmiJY{
  Y.~Shadmi and Y.~Shirman,
  ``Dynamical supersymmetry breaking,''
  Rev.\ Mod.\ Phys.\  {\bf 72}, 25 (2000)
  [arXiv:hep-th/9907225].
}

\lref\IntriligatorDD{
  K.~Intriligator, N.~Seiberg and D.~Shih,
  ``Dynamical SUSY breaking in meta-stable vacua,''
  JHEP {\bf 0604}, 021 (2006)
  [arXiv:hep-th/0602239].
}

\lref\EllisVI{
  J.~R.~Ellis, C.~H.~Llewellyn Smith and G.~G.~Ross,
  ``Will The Universe Become Supersymmetric?,''
  Phys.\ Lett.\  B {\bf 114}, 227 (1982).
}

\lref\DimopoulosWW{
  S.~Dimopoulos, G.~R.~Dvali, R.~Rattazzi and G.~F.~Giudice,
  ``Dynamical soft terms with unbroken supersymmetry,''
  Nucl.\ Phys.\  B {\bf 510}, 12 (1998)
  [arXiv:hep-ph/9705307].
}

\lref\DineAG{
 M.~Dine, A.~E.~Nelson, Y.~Nir and Y.~Shirman,
 ``New tools for low-energy dynamical supersymmetry breaking,''
 Phys.\ Rev.\ D {\bf 53}, 2658 (1996)
 [arXiv:hep-ph/9507378].
}

\lref\LutyVR{
  M.~A.~Luty and J.~Terning,
  ``Improved single sector supersymmetry breaking,''
  Phys.\ Rev.\  D {\bf 62}, 075006 (2000)
  [arXiv:hep-ph/9812290].
}

\lref\BanksDF{
  T.~Banks,
  ``Cosmological supersymmetry breaking and the power of the pentagon: A  model
  of low energy particle physics,''
  arXiv:hep-ph/0510159.
}

\lref\AffleckXZ{
  I.~Affleck, M.~Dine and N.~Seiberg,
  ``Dynamical Supersymmetry Breaking In Four-Dimensions And Its
  Phenomenological Implications,''
  Nucl.\ Phys.\ B {\bf 256}, 557 (1985).
}

\lref\KribsAC{
  G.~D.~Kribs, E.~Poppitz and N.~Weiner,
  ``Flavor in Supersymmetry with an Extended R-symmetry,''
  arXiv:0712.2039 [hep-ph].
}

\lref\ShihAV{
  D.~Shih,
  ``Spontaneous R-symmetry breaking in O'Raifeartaigh models,''
  JHEP {\bf 0802}, 091 (2008)
  [arXiv:hep-th/0703196].
}

\lref\CheungES{
  C.~Cheung, A.~L.~Fitzpatrick and D.~Shih,
  ``(Extra)Ordinary Gauge Mediation,''
  arXiv:0710.3585 [hep-ph].
}

\lref\NappiHM{
  C.~R.~Nappi and B.~A.~Ovrut,
  ``Supersymmetric Extension Of The SU(3) X SU(2) X U(1) Model,''
  Phys.\ Lett.\  B {\bf 113}, 175 (1982).
}

\lref\DineXT{
  M.~Dine and J.~Mason,
  ``Gauge mediation in metastable vacua,''
  Phys.\ Rev.\  D {\bf 77}, 016005 (2008)
  [arXiv:hep-ph/0611312].
}

\lref\MurayamaYF{
  H.~Murayama and Y.~Nomura,
  ``Gauge mediation simplified,''
  Phys.\ Rev.\ Lett.\  {\bf 98}, 151803 (2007)
  [arXiv:hep-ph/0612186].
}

\lref\AharonyMY{
  O.~Aharony and N.~Seiberg,
  ``Naturalized and simplified gauge mediation,''
  JHEP {\bf 0702}, 054 (2007)
  [arXiv:hep-ph/0612308].
}

\lref\CsakiWI{
  C.~Csaki, Y.~Shirman and J.~Terning,
  ``A simple model of low-scale direct gauge mediation,''
  JHEP {\bf 0705}, 099 (2007)
  [arXiv:hep-ph/0612241].
}

\lref\IntriligatorPY{
  K.~Intriligator, N.~Seiberg and D.~Shih,
  ``Supersymmetry Breaking, R-Symmetry Breaking and Metastable Vacua,''
  JHEP {\bf 0707}, 017 (2007)
  [arXiv:hep-th/0703281].
}

\lref\FrancoES{
  S.~Franco and A.~M.~Uranga,
  ``Dynamical SUSY breaking at meta-stable minima from D-branes at obstructed
  geometries,''
  JHEP {\bf 0606}, 031 (2006)
  [arXiv:hep-th/0604136].
}

\lref\DimopoulosIG{
  S.~Dimopoulos and G.~F.~Giudice,
  ``Multi-messenger theories of gauge-mediated supersymmetry breaking,''
  Phys.\ Lett.\  B {\bf 393}, 72 (1997)
  [arXiv:hep-ph/9609344].
}

\lref\OoguriPJ{
  H.~Ooguri and Y.~Ookouchi,
  ``Landscape of supersymmetry breaking vacua in geometrically realized gauge
  theories,''
  Nucl.\ Phys.\  B {\bf 755}, 239 (2006)
  [arXiv:hep-th/0606061].
}

\lref\AmaritiVK{
  A.~Amariti, L.~Girardello and A.~Mariotti,
  ``Non-supersymmetric meta-stable vacua in SU(N) SQCD with adjoint matter,''
  JHEP {\bf 0612}, 058 (2006)
  [arXiv:hep-th/0608063].
}

\lref\GiveonWP{
  A.~Giveon, A.~Katz and Z.~Komargodski,
  ``On SQCD with massive and massless flavors,''
  JHEP {\bf 0806}, 003 (2008)
  [arXiv:0804.1805 [hep-th]].
}

\lref\ChackoUU{
  Z.~Chacko, M.~A.~Luty and E.~Ponton,
  ``Dynamical determination of the unification scale by gauge-mediated
  supersymmetry breaking,''
  Phys.\ Rev.\  D {\bf 59}, 035004 (1999)
  [arXiv:hep-ph/9806398].
}

\lref\DineVC{
 M.~Dine, A.~E.~Nelson and Y.~Shirman,
 ``Low-Energy Dynamical Supersymmetry Breaking Simplified,''
 Phys.\ Rev.\ D {\bf 51}, 1362 (1995)
 [arXiv:hep-ph/9408384].
}

\lref\ShadmiMD{
 Y.~Shadmi,
 ``Gauge-mediated supersymmetry breaking without fundamental singlets,''
 Phys.\ Lett.\ B {\bf 405}, 99 (1997)
 [arXiv:hep-ph/9703312].
}

\lref\DineGM{
  M.~Dine, J.~L.~Feng and E.~Silverstein,
  ``Retrofitting O'Raifeartaigh models with dynamical scales,''
  Phys.\ Rev.\  D {\bf 74}, 095012 (2006)
  [arXiv:hep-th/0608159].
}

\lref\KraussZC{
  L.~M.~Krauss and F.~Wilczek,
  ``Discrete Gauge Symmetry in Continuum Theories,''
  Phys.\ Rev.\ Lett.\  {\bf 62}, 1221 (1989).
}

\lref\ISSiii{
  K.~Intriligator, D.~Shih and M.~Sudano, ``Surveying Pseudomoduli: the Good, the Bad and the Incalculable," in preparation.
  }

\lref\GiudiceNI{
  G.~F.~Giudice and R.~Rattazzi,
  ``Extracting supersymmetry-breaking effects from wave-function
  renormalization,''
  Nucl.\ Phys.\  B {\bf 511}, 25 (1998)
  [arXiv:hep-ph/9706540].
}

\lref\AGLR{
  N.~Arkani-Hamed, G.~F.~Giudice, M.~A.~Luty and R.~Rattazzi,
  ``Supersymmetry-breaking loops from analytic continuation into  superspace,''
  Phys.\ Rev.\  D {\bf 58}, 115005 (1998)
  [arXiv:hep-ph/9803290].
}

\lref\SeibergPQ{
  N.~Seiberg,
  ``Electric - magnetic duality in supersymmetric nonAbelian gauge theories,''
  Nucl.\ Phys.\  B {\bf 435}, 129 (1995)
  [arXiv:hep-th/9411149].
}

\lref\BaggerHH{
  J.~Bagger, E.~Poppitz and L.~Randall,
  ``The R axion from dynamical supersymmetry breaking,''
  Nucl.\ Phys.\  B {\bf 426}, 3 (1994)
  [arXiv:hep-ph/9405345].
}

\lref\IbanezHV{
  L.~E.~Ibanez and G.~G.~Ross,
  ``Discrete gauge symmetry anomalies,''
  Phys.\ Lett.\  B {\bf 260}, 291 (1991).
}

\lref\IbanezJI{
  L.~E.~Ibanez,
  ``More about discrete gauge anomalies,''
  Nucl.\ Phys.\  B {\bf 398}, 301 (1993)
  [arXiv:hep-ph/9210211].
}

\lref\GiudiceCA{
  G.~F.~Giudice, H.~D.~Kim and R.~Rattazzi,
  ``Natural mu and Bmu in gauge mediation,''
  Phys.\ Lett.\  B {\bf 660}, 545 (2008)
  [arXiv:0711.4448 [hep-ph]].
}

\lref\DimopoulosGY{
  S.~Dimopoulos, G.~F.~Giudice and A.~Pomarol,
  ``Dark matter in theories of gauge-mediated supersymmetry breaking,''
  Phys.\ Lett.\  B {\bf 389}, 37 (1996)
  [arXiv:hep-ph/9607225].
}

\lref\FengKE{
  J.~L.~Feng, C.~G.~Lester, Y.~Nir and Y.~Shadmi,
  ``The Standard Model and Supersymmetric Flavor Puzzles at the Large Hadron
  Collider,''
  arXiv:0712.0674 [hep-ph].
}

\lref\GiveonEF{
  A.~Giveon and D.~Kutasov,
  ``Stable and Metastable Vacua in SQCD,''
  Nucl.\ Phys.\  B {\bf 796}, 25 (2008)
  [arXiv:0710.0894 [hep-th]].
}

\lref\KitanoWM{
  R.~Kitano,
  ``Dynamical GUT breaking and mu-term driven supersymmetry breaking,''
  Phys.\ Rev.\  D {\bf 74}, 115002 (2006)
  [arXiv:hep-ph/0606129].
}

\lref\KitanoWZ{
  R.~Kitano,
  ``Gravitational gauge mediation,''
  Phys.\ Lett.\  B {\bf 641}, 203 (2006)
  [arXiv:hep-ph/0607090].
}

\lref\AmaritiVK{
  A.~Amariti, L.~Girardello and A.~Mariotti,
  ``Non-supersymmetric meta-stable vacua in SU(N) SQCD with adjoint matter,''
  JHEP {\bf 0612}, 058 (2006)
  [arXiv:hep-th/0608063].
}

\lref\DineXT{
  M.~Dine and J.~Mason,
  ``Gauge mediation in metastable vacua,''
  Phys.\ Rev.\  D {\bf 77}, 016005 (2008)
  [arXiv:hep-ph/0611312].
}

\lref\KitanoXG{
  R.~Kitano, H.~Ooguri and Y.~Ookouchi,
  ``Direct mediation of meta-stable supersymmetry breaking,''
  Phys.\ Rev.\  D {\bf 75}, 045022 (2007)
  [arXiv:hep-ph/0612139].
}

\lref\AmaritiQU{
  A.~Amariti, L.~Girardello and A.~Mariotti,
  ``On meta-stable SQCD with adjoint matter and gauge mediation,''
  Fortsch.\ Phys.\  {\bf 55}, 627 (2007)
  [arXiv:hep-th/0701121].
}

\lref\AbelJX{
  S.~Abel, C.~Durnford, J.~Jaeckel and V.~V.~Khoze,
  ``Dynamical breaking of $U(1)_{R}$ and supersymmetry in a metastable vacuum,''
  Phys.\ Lett.\  B {\bf 661}, 201 (2008)
  [arXiv:0707.2958 [hep-ph]].
}

\lref\HabaRJ{
  N.~Haba and N.~Maru,
  ``A Simple Model of Direct Gauge Mediation of Metastable Supersymmetry
  Breaking,''
  Phys.\ Rev.\  D {\bf 76}, 115019 (2007)
  [arXiv:0709.2945 [hep-ph]].
}

\lref\AbelNR{
  S.~A.~Abel, C.~Durnford, J.~Jaeckel and V.~V.~Khoze,
  ``Patterns of Gauge Mediation in Metastable SUSY Breaking,''
  JHEP {\bf 0802}, 074 (2008)
  [arXiv:0712.1812 [hep-ph]].
}

\lref\ZurZG{
  B.~K.~Zur, L.~Mazzucato and Y.~Oz,
  ``Direct Mediation and a Visible Metastable Supersymmetry Breaking Sector,''
  arXiv:0807.4543 [hep-ph].
}

\lref\NelsonNF{
  A.~E.~Nelson and N.~Seiberg,
  ``R symmetry breaking versus supersymmetry breaking,''
  Nucl.\ Phys.\  B {\bf 416}, 46 (1994)
  [arXiv:hep-ph/9309299].
}


\Title{\vbox{\baselineskip12pt \hbox{}}} {\vbox{\centerline{Dynamical SUSY and R-Symmetry Breaking
in}\vskip5pt\centerline{SQCD  with Massive and Massless Flavors}}}
\smallskip
\centerline{Amit Giveon,$^{1}$ Andrey Katz,$^2$ Zohar Komargodski$^3$ and David Shih$^4$}
\smallskip
\bigskip
\centerline{$^1${\it Racah Institute of Physics, The Hebrew University, Jerusalem 91904, Israel}}
\medskip
\centerline{$^2${\it Physics Department, Technion-Israel Institute of Technology, Haifa 32000, Israel}}
\medskip
\centerline{$^3${\it Department of Particle Physics, The Weizmann Institute of Science, Rehovot 76100, Israel }}
\medskip
\centerline{$^4${\it School of Natural Sciences, Institute for
Advanced Study, Princeton, NJ 08540 USA}}
\vskip 1cm

\noindent

We show that supersymmetry and R-symmetry can be dynamically broken in a long-lived metastable vacuum of SQCD
with massive and massless flavors. The vacuum results from a competition of a (leading) two-loop effect and
small ``Planck" suppressed higher-dimension operators. This mechanism provides a particularly simple realization
of dynamical SUSY and R-symmetry breaking, and as such it is a good starting point for building
phenomenologically viable models of gauge mediation. We take a preliminary step in this direction by
constructing a complete model of minimal gauge mediation. Here we find that the parameters of the model are
surprisingly constrained by the hidden sector. Similar mechanisms for creating long-lived states operate in a
large class of models.

\bigskip
\bigskip

\Date{July 2008}

\newsec{Introduction}

Dynamical supersymmetry breaking (DSB) via strong gauge dynamics has long been an attractive mechanism for
explaining the large hierarchy between the weak scale and the Planck scale~\WittenNF. Early, seminal models of
DSB were important ``existence proofs" -- however, they tended to be rather complicated (see
e.g.~\refs{\GiudiceBP\PoppitzVD\TerningTH-\ShadmiJY} for reviews and references). This was tied to the
assumption that the DSB vacuum had to be the global minimum of the potential. By relaxing this assumption and
allowing for metastability, much simpler models of DSB have recently been constructed, starting with the
discovery of metastable vacua in SQCD with massive flavors~\IntriligatorDD. (Of course, metastability and its
connection to SUSY breaking are not new ideas; for earlier works see
e.g.~\refs{\EllisVI\DineAG\LutyVR\DimopoulosWW -\BanksDF}.) This work has opened new model building avenues,
especially for models of direct gauge mediation. Since metastable SUSY-breaking models can be vector-like, they
can naturally have large flavor symmetries. This makes them more amenable for direct gauge mediation, where a
subgroup of the flavor symmetry is gauged and identified with the SM gauge group~\AffleckXZ.

Unfortunately, despite all its positive features, the metastable
vacuum of SQCD with massive flavors has one major phenomenological
defect: it preserves a continuous $U(1)_R$ symmetry. Without
breaking this symmetry somehow, the gauginos of the visible sector
would remain massless.\foot{Two comments. First, the R-symmetry is
broken explicitly by nonperturbative effects but this is generally
too small to allow for realistic gaugino masses. Second, we are
assuming here standard Majorana gaugino masses. An interesting
alternative whose phenomenology has recently been revisited in \KribsAC\ is to have Dirac
gaugino masses, in which case the R-symmetry need not be broken at
all.}

Various approaches have been tried to get around this problem. Motivated by the fact that the low-energy
effective theory of massive SQCD is simply a generalized O'Raifeartaigh model, more sophisticated O'Raifeartaigh
models with R-charge assignments different from $0$ or $2$ were studied in~\ShihAV, and it was shown that such
models could lead to spontaneous R-symmetry breaking. A detailed study of the phenomenology of such models was
undertaken in~\CheungES; however, no UV complete dynamical realization of these models is known. (Non-renormalizable realizations based on massive SQCD with baryonic deformations were constructed and studied in \refs{\AbelJX,\AbelNR}.)  Alternatively, one can
gauge an auxiliary $U(1)$ symmetry of the problem; the competition between the gauge and Yukawa couplings can
lead to a spontaneous R-symmetry breaking minimum of the one-loop effective
potential~\refs{\NappiHM\DineXT-\CsakiWI}. However, here there generally has to be some fine tuning of the
couplings in order to achieve the desired R-symmetry breaking~\IntriligatorPY. Finally, a third (and perhaps
most obvious) possibility is to break the approximate R-symmetry explicitly, but  on general grounds \NelsonNF\ this generically leads to
proximal SUSY vacua and some tuning has to be invoked (see e.g.\ \refs{\KitanoXG\HabaRJ-\ZurZG} for models). One possibility that has been explored is to have
explicit R-symmetry breaking only through Planck-suppressed couplings to
messengers~\refs{\MurayamaYF,\AharonyMY}. However, in this case the messenger scale and the messenger SUSY
splittings come from different sources, so there are in general extra CP phases in the gaugino masses when
doublet/triplet splitting of the messengers is taken into account. Similarly, messenger-parity is not guaranteed
and there is a danger of inducing spontaneous breaking of the SM gauge group~\DimopoulosIG.

In this work we present another mechanism of spontaneous R-symmetry breaking which can be viewed as an
improvement upon previous attempts, inasmuch as it is simple, more natural, dynamical and phenomenologically
viable. The crucial observation is that all the abovementioned attempts at spontaneous R-symmetry breaking
assumed that the R-charged pseudo-moduli obtain their potential at one-loop. In some instances, however,
pseudo-moduli do not obtain their potentials until two-loops, and then the story can be quite different. Here,
we will study one of the simplest illustrations of this phenomenon: SQCD with both massive and massless flavors.
In this model, the meson composed of the massless quarks remains massless at one-loop in the low-energy
effective theory~\FrancoES. Much above the strong coupling scale the theory exhibits nonperturbative runaway
behavior; below the strong coupling scale, the two-loop potential was recently computed in the low-energy
effective theory and was shown to be a monotonically decreasing function of the meson VEV~\GiveonWP. Thus the
model probably has runaway for all values of the field and has no vacuum at all.

In this paper we show how the runaway is automatically stabilized by ``Planck" suppressed higher-dimension
operators in the superpotential. The result is a local vacuum far from the origin but well below the cutoff scale of the
effective low-energy theory, where both SUSY and R-symmetry are spontaneously broken.\foot{A similar idea in a
different context was proposed in~\ChackoUU\ to explain the hierarchy between the Planck scale and the GUT
scale. See also \refs{\FrancoES,\OoguriPJ,\AmaritiVK} for recent models based on massive SQCD where dangerous pseudo-moduli directions were stabilized with higher-dimension operators. However in these
latter examples the pseudo-moduli in question were stabilized at the origin and were not the sources of
R-symmetry breaking.}

Aside from the SUSY-breaking minimum, the higher-dimension operators also restore a supersymmetric solution. We
show that the distance between the supersymmetric and SUSY-breaking solution is parametrically large, ensuring
that the metastable vacuum is parametrically long-lived.

Consequently, SQCD with massive and massless flavors has all the necessary ingredients, including large
spontaneous R-symmetry breaking, for SUSY model building. We illustrate this by writing down the simplest
application of our model, namely a complete model of ``minimal" gauge mediation~\refs{\DineAG,\DineVC}. The role
of the singlet coupling to messengers in our model is played by the meson composed of the massless electric
quarks. As in~\ShadmiMD, we couple the meson to the messengers through non-renormalizable interactions in the
UV. We find that if the runaway is stabilized by a quartic coupling in the UV, surprisingly stringent
constraints on such models of minimal gauge mediation come from the requirements of calculability in the
SUSY-breaking sector together with non-tachyonic messengers.\foot{These constraints can be relaxed if one
chooses to stabilize the runaway by even higher-dimension operators. The main drawback of this possibility is
that a more complicated discrete symmetry is needed to maintain such a structure.} As we show further this model
can be successfully retrofitted~\DineGM\ and all the couplings can be explained in the context of discrete gauge
symmetries~\KraussZC.

Of course, the real promise of our new mechanism of dynamical SUSY and R-symmetry breaking is in its potential
applications to models of direct gauge mediation, where there is no invariant distinction between messenger and
SUSY-breaking fields. Since SQCD with massive and massless flavors is vector-like, it automatically comes
equipped with large global flavor symmetries. It is easy to imagine gauging a subgroup of these flavor
symmetries and identifying it with the SM gauge group. Then SUSY breaking will be automatically communicated to
the SM via gauge mediation, and unlike in~\IntriligatorDD, the MSSM gauginos will acquire viable soft masses. In
this context one still has to overcome the long-standing challenge of the Landau pole problem, since there will
generally be lots of extra matter charged under the SM gauge group. This is a very interesting problem that we
will reserve for future work.

Our paper is organized as follows. In section~2 we review SQCD with massive and massless flavors. We pay special
attention to the theory far out along the pseudo-moduli space (where the runaway will ultimately be stabilized).
In this regime, the theory can be described by an effective supersymmetric Lagrangian, and the leading-log
approximation to the runaway potential has been calculated as part of a more general analysis in~\ISSiii, using
the techniques of wavefunction renormalization \refs{\GiudiceNI,\AGLR}. However, to keep the discussion in this
paper self-contained we present a sketch of the calculation in the appendix. In section~3 we introduce the
stabilization of the two-loop runaway potential via higher-dimension operators, discuss the spectrum of
particles around the SUSY-breaking vacuum, and derive constraints on the parameters of the model which follow
from calculability and longevity. Section~4 is dedicated to constructing and analyzing a model of minimal gauge
mediation using massive+massless SQCD as the SUSY-breaking sector.


\newsec{SQCD with Massive and Massless Flavors}

\subsec{An overview of the model}

In this section we review the main results of~\GiveonWP, while simultaneously establishing our notation and
conventions (which differ slightly from those of~\GiveonWP). We will consider SUSY QCD with $SU(N_c)$ gauge
group and $N_{f}$ flavors in the free magnetic phase ($N_c<N_f<3N_c/2$). We will take $N_{f0}<N_c$ of the
flavors to be massless with the other $N_{f1}=N_f-N_{f0}$ having equal mass.\foot{It is trivial to ensure the
naturalness of this structure by imposing appropriate (discrete) symmetries which allow mass terms for some of
the quarks but not others. We will have more to say about this in the next sections.} When all the flavors are
massive (but light compared to the strong coupling scale $\Lambda$), this model exhibits a metastable
supersymmetry breaking state~\IntriligatorDD. However, when some of the flavors are massless, the situation is
very different: some of the pseudo-moduli of the low-energy effective theory are not stabilized at one-loop
~\FrancoES. Instead, they acquire a potential at two-loops and it was recently shown in~\GiveonWP\ that this
potential is monotonically decreasing along the pseudo-moduli space, leading to runaway behavior.

Now let us describe the low-energy effective theory in more detail. Using Seiberg duality~\SeibergPQ, we have a
weakly coupled description at low energies in terms of an IR-free $SU(N)$ gauge theory (with $N=N_f-N_c$). The
matter content consists of a gauge singlet $N_{f}\times N_f$ meson matrix $\Phi$ and $N_f$ flavors of magnetic
quarks $\varphi$, $\tilde \varphi$.  It will be convenient in the following to split the $N_f$ flavor indices
into $N_{f0}$ and $N_{f1}$ sized blocks, i.e.\ \eqn\convfields{ \Phi = \pmatrix{ \Phi_{11} &
\Phi_{10}\cr\Phi_{01} & \Phi_{00}}~,\qquad \varphi=\pmatrix{\varphi_1\cr \varphi_0}~,\qquad \tilde
\varphi^T=\pmatrix{\tilde \varphi_1\cr\tilde \varphi_0}~, } with $\Phi_{ij}$ an $N_{fi}\times N_{fj}$ matrix and
$\varphi_i$, $\tilde \varphi_i^T$ $N_{fi}\times N$ matrices. The fields can be normalized to have canonical
K\"ahler potential plus uncalculable higher-dimensional corrections suppressed by powers of $\Lambda$. In this
normalization, the superpotential of the theory is \eqn\sqcd{
    W=h\Tr\,\Phi_{ij} \varphi_j\tilde\varphi_i-h\mu^2\Tr\,\Phi_{11}~,
     }
where $h$ is an $\CO(1)$ coupling and the traces are taken over the uncontracted flavor indices of the different
fields. In what follows we will assume without loss of generality that all the couplings are real and positive.
Note that although the global symmetry is no longer $SU(N_f)$, we are still denoting all the Yukawa couplings by
the same symbol $h$ -- at the scale $\Lambda$ this is the case up to corrections which vanish as
${\mu\over\Lambda}\to 0$.
Of course, below the scale $\Lambda$ these couplings will run differently, and we will take this into account
when necessary.

The representations of the fields under all the gauge and global symmetries preserved by the
superpotential~\sqcd\ are listed in the table below. Note that there are four independent $U(1)$ symmetries
(including the R-symmetry), neglecting anomalies as in~\IntriligatorDD. \eqn\globalsym{
 \matrix{
                        & SU(N) & \Big[SU(N_{f0})_L & SU(N_{f0})_R & SU(N_{f1}) & U(1)_B & U(1)_1 & U(1)_2 & U(1)_R\Big] \cr
          & \cr
         \Phi_{11} & {\bf 1} & {\bf 1}  & {\bf 1} &  {\rm {\bf adj}}\oplus {\bf 1} &   0 & 0 &  0 & 2\cr
         \Phi_{10} & {\bf 1}  &  \square & {\bf 1} & \overline{\square}  &  0 & -1 &  1 & {3/2} \cr
         \Phi_{01} & {\bf 1}  & {\bf 1} & \overline{\square} & \square    &  0 & 1 &  0 & {3/2}\cr
         \Phi_{00} & {\bf 1}  & \square & \overline{\square}  &  {\bf 1} &  0  & 0 &  1 & 1 \cr
         \varphi_1 & \square & {\bf 1} &  {\bf 1} & \overline{\square} & 1 & 0 & 0 & 0 \cr
         \varphi_0 & \square &  \overline{\square} & {\bf 1} & {\bf 1} & 1 & 1 & -1  & {1/2} \cr
         \tilde\varphi_1 & \overline{\square} & {\bf 1} & {\bf 1} & \square & -1 & 0 & 0  & 0\cr
         \tilde\varphi_0 & \overline{\square} &  {\bf 1} & \square & {\bf 1} & -1 & -1 & 0 &{1/2} \cr
         }
}
The reason for the funny $U(1)$ charges will become apparent in the next section.

Since $N_{f1}=N_f-N_{f0}>N_f-N_c$, rank conditions mean that SUSY is spontaneously broken at tree-level, as
in~\IntriligatorDD. The tree-level scalar potential is minimized along the following pseudo-moduli space:
 \eqn\solu{\eqalign{
  &\Phi_{11}=\pmatrix{0 & 0 \cr 0 & X}~,\qquad \Phi_{10}=\pmatrix{ 0 \cr Y}~,\qquad  \varphi_1=\left(\matrix{\chi
 \cr 0}\right)~,\qquad \tilde \varphi_1^T =\left(\matrix{\tilde\chi \cr 0}\right)~,\qquad \chi\tilde\chi=\mu^2\unit_N~,\cr
  &\Phi_{01}=\pmatrix{0 &\,\, \tilde Y}~,\qquad\,\,
 {\rm arbitrary}\ \Phi_{00}~,\qquad \varphi_0=\tilde\varphi_0=0~,
\cr
}}
with vacuum energy
\eqn\vace{
 V_0 = (N_{f1}-N)h^2\mu^4~.
 }
Here $\chi$, $\tilde\chi$ are $N\times N$ matrices, $X$ is an $(N_{f1}-N)\times (N_{f1}-N)$ matrix, $Y$ and
$\tilde Y^T$ are $(N_{f1}-N)\times N_{f0}$ matrices, and $\Phi_{00}$ is an $N_{f0}\times N_{f0}$ matrix. At the
origin of the pseudo-moduli space~\solu, the entire gauge symmetry and part of the global symmetry are broken:
\eqn\unbrokensym{
 SU(N)\times \Big[SU(N_{f1})\times U(1)_B\Big] \to \Big[SU(N)_D\times SU(N_{f1}-N)\times  U(1)_B'\Big]~,
 }
with the other global symmetries remaining unaffected.

As mentioned at the beginning of this section, the main novelty introduced by the massless quarks is the fact
that at one-loop, only some of pseudo-moduli are lifted by the Coleman-Weinberg potential~\FrancoES.
Specifically, all the fields except $\Phi_{00}$ are lifted and stabilized at the origin by the one-loop
potential. So to understand the dynamics of this model we are driven to a two-loop computation on the
pseudo-moduli space parameterized by $\Phi_{00}$. This has been considered in detail in~\GiveonWP. There it was
found that the origin of field space is {\it destabilized}, in other words, $\Phi_{00}$ acquires a negative mass
squared term around the origin \eqn\mass{
    V_{eff}= -h^2\mu^2\left({\alpha_h\over 4\pi}\right)^2N(N_{c}-N_{f0})\left(1+{\pi^2\over 6}-\log 4\right)\Tr(\Phi_{00}^\dagger
\Phi_{00})+\CO(\Phi_{00}^4)~, } where $\alpha_h\equiv{h^2\over 4\pi}$. Moreover, it was shown in~\GiveonWP\ that
the two-loop potential for $\Phi_{00}$ is monotonically decreasing, leading to runaway behavior. This result has
important physical implications for ISS based model building and it can be interpreted geometrically in the
appropriate brane configuration~\GiveonWP. In the next subsection we will study the model in more detail in the
$\langle\Phi_{00}\rangle\gg \mu$ regime.

\subsec{Effective theory at large $\langle\Phi_{00}\rangle$}

Since the pseudo-modulus $\Phi_{00}$ has a runaway potential, we are motivated to study the theory in the large
$\Phi_{00}\gg\mu$ regime. There the analysis of the pseudo-modulus potential simplifies considerably -- we can
use the wavefunction renormalization techniques of~\refs{\GiudiceNI,\AGLR} to compute the two-loop potential in
the leading-log approximation using only one-loop anomalous dimensions.

To be more precise, we are interested in the theory for generic $\langle\Phi_{00}\rangle$ in the regime
 \eqn\inequal{
  h\mu\ll h \langle\Phi_{00}\rangle\ll \Lambda~.
  }
The point is that below the scale $h\langle\Phi_{00}\rangle$ we can integrate out supersymmetrically the
$N_{f0}$ magnetic quarks coupling to $\Phi_{00}$. Then we are left with an effective supersymmetric theory of
the form \eqn\efftheory{\eqalign{
  & K_{eff} = \CZ_{\Phi_{ij}}\Tr\,\Phi_{ij}^\dagger\Phi_{ij} + \CZ_{\varphi_1}(\Tr\,\varphi_1^\dagger \varphi_1 + \Tr\,\tilde\varphi_1^\dagger \tilde\varphi_1)+  \dots~, \cr
 & W_{eff} = h\Tr\,\Phi_{11}\varphi_1\tilde \varphi_1-h\mu^2\Tr\,\Phi_{11}
 -h\Tr\,\tilde\varphi_1\Phi_{10}\Phi_{00}^{-1}\Phi_{01}\varphi_1~,
}} where the wavefunction factors in the K\"ahler potential are functions of $\Phi_{00}$ and where $\dots$
includes both the uncalculable $\CO(1/\Lambda)$ corrections from the duality together with the calculable
$\CO(1/\langle\Phi_{00}\rangle)$ corrections from integrating out $\varphi_0$, $\tilde\varphi_0$.

The effective theory still has an $SU(N)$ gauge symmetry; the gauge coupling can be either IR or asymptotically
free depending on whether $N_{f1}>3N$ or $N_{f1}<3N$, respectively. The latter would be problematic because then
the K\"ahler potential would no longer be well-behaved around the origin where we need to compute loops to
understand the SUSY-breaking vacuum. Thus in order to have a reliable SUSY-breaking vacuum we must require
\eqn\Nfreq{ N_{f1} > 3N=3(N_f-N_c)~. } This goes beyond the usual free-magnetic phase requirement $N_f>3N$. Note
that \Nfreq\ is equivalent to \eqn\Nfreqii{ N_{f0} < 3N_c-2N_f < N_c~. }

The effective theory is essentially the ``macroscopic model" of~\IntriligatorDD\ but with a wavefunction factor
in the K\"ahler potential that depends on $\Phi_{00}$. The theory still breaks SUSY via the rank condition and
stabilizes all the pseudo-moduli. The vacuum energy is: \eqn\effthvace{ V_0 =
(N_{f1}-N)h^2\mu^4\CZ_{\Phi_{11}}^{-1}~. } This can be interpreted as the effective potential for $\Phi_{00}$,
after using the RGEs to compute $\CZ_{\Phi_{11}}$ in the leading-log approximation. The computation of
$\CZ_{\Phi_{11}}$ has been performed in~\ISSiii\ as a special case of a more general survey of pseudo-moduli.
Here we simply quote the answer;  we will reproduce the calculation in appendix A for the sake of completeness.
\eqn\wvfnPhidepiii{\eqalign{ \log \CZ_{\Phi_{11}} &=N\left({\alpha_h\over 4\pi}\right)^2
\Tr\left(\log{\Phi_{00}^\dagger\Phi_{00}\over \mu^2}\right)^2 +\dots~,\cr }} where $\dots$ are subleading in the
two-loop leading-log approximation. Plugging this into~\effthvace, we have in the leading-log approximation
\eqn\asymp{
    V_{eff}=-(N_{f1}-N)Nh^2\mu^4\left({\alpha_{h}\over 4\pi}\right)^2\Tr\,\left(\log{\Phi_{00}^\dagger\Phi_{00}\over
    \mu^2}\right)^2~.
    }
Of course,~\asymp\ can also be computed more directly by taking the large field limit of the two-loop potential
computed in~\GiveonWP.


\newsec{Stabilizing the Runaway}

\subsec{Adding a small quartic term}

In this section, we will show how the runaway can be turned into a metastable SUSY-breaking vacuum by allowing for generic non-renormalizable superpotential interactions. One can easily imagine that such terms are present at the Planck scale, generated by the theory of quantum gravity. Alternatively, they could be trivially generated at an intermediate scale by integrating out massive matter in some renormalizable UV completion. In any event, the point is that, since the runaway direction is a dimension two meson in the electric theory, the lowest-dimension non-renormalizable operators -- dimension four interactions amongst the electric quarks -- become mass terms for the runaway fields in the IR. So the runaway is automatically stabilized by generic higher-dimensional superpotential interactions.

Instead of considering the most general quartic terms for the electric
quarks, we will focus only those involving the $N_{f0}$ massless quarks; these become in the IR:
 \eqn\quarticdefii{ \delta W = {1\over2}
m_{f_1g_1f_2g_2}(\Phi_{00})_{f_1g_1}(\Phi_{00})_{f_2g_2}~.} It would be interesting to study in detail more
general deformations involving $\Phi_{01}$, $\Phi_{10}$ and $\Phi_{11}$, but we will not do so in this paper. We
will briefly discuss the possible effects of such deformations at the end of the next subsection. The upshot is
that, while such deformations may complicate the details of our analysis, it is not obvious that they will
change the qualitative conclusions.

For $N_{f0}>2$, the quartic deformations in \quarticdefii\
necessarily break the $SU(N_{f0})_L\times SU(N_{f0})_R$ symmetry down to a subgroup. For simplicity, we will
choose to preserve $SU(N_{f0})_D$. We will further simplify the analysis by considering only one of the two
independent deformations that respect this symmetry: \eqn\quarticdef{
 \delta W = {1\over2}h \epsilon \mu\,\Tr\,\Phi_{00}^2~.
 }
One could of course consider an arbitrary mixture of~\quarticdef\ and the other $SU(N_{f0})_D$ symmetric
deformation $(\Tr\,\Phi_{00})^2$. However, since this complication will have no effect on the essential physics
that we will discuss below, we will ignore it. Notice that in~\quarticdef\ we have written the deformation
parameter in units of $h\mu$, with a new dimensionless parameter $\epsilon$. This will simplify the notation
throughout.

In order for the deformation to not overwhelm the negative two-loop mass-squared for $\Phi_{00}$ \mass\ around
the origin, we need $\epsilon$ to be a small parameter (as its name suggests), specifically
\eqn\reqmi{
 \epsilon \lesssim {\alpha_h\over 4\pi}~.
 }
Ordinarily this small parameter would seem rather artificial; however, here the UV completion can make it
natural -- all we need is that the UV scale that suppresses~\quarticdef\ in the electric theory is sufficiently
large compared to $\Lambda$.

We are interested in whether the deformation can stabilize the runaway at large $\Phi_{00}\gg \mu$. In this
regime, the effective theory~\efftheory\ deformed by~\quarticdefii\ is a valid description. Thus the vacuum
energy along the pseudo-moduli space $\Phi_{00}$ is
 \eqn\defvace{ V_0 =  (N_{f1}-N)h^2\mu^4 \CZ_{\Phi_{11}}^{-1} +
h^2\epsilon^2\mu^2\Tr\,\Phi_{00}^\dagger\Phi_{00}~. } Here we are neglecting the wavefunction renormalization of
$\Phi_{00}$ itself -- since we are going to be balancing the two terms in~\defvace\ against each other to find
the metastable vacuum, any corrections from $\CZ_{\Phi_{00}}$ are obviously going to be subleading.

Substituting the calculation \asymp\ of the previous section, it is straightforward to minimize the potential,
\eqn\defvacesub{ V_{eff} = -(N_{f1}-N)Nh^2\mu^4\left({\alpha_{h}\over
4\pi}\right)^2\Tr\,\left(\log{\Phi_{00}^\dagger\Phi_{00}\over
    \mu^2}\right)^2 + h^2\epsilon^2\mu^2\Tr\,\Phi_{00}^\dagger\Phi_{00}~,
    }
and find the local minimum at
\eqn\msvac{
  \langle\Phi_{00}\rangle\approx {\mu\over\delta} \sqrt{\log {1\over\delta}}\,\,\unit_{N_{f0}}~,\qquad {1\over\delta}\equiv  \sqrt{(N_{f1}-N)N}\left({\alpha_h\over 2\pi}\right){1\over\epsilon}~.
 }
The corrections to~\msvac\ are of the form $\langle\Phi_{00}\rangle\to\langle\Phi_{00}\rangle\times\left(1+
\CO\left({1\over\log\delta},\,{\log(\log\delta)\over\log\delta}\right)\right)$ and they come not only from
minimizing the potential~\defvacesub, but also from the subleading-log corrections to the wavefunction
$\CZ_{\Phi_{11}}$. It is very important (and fortunate) for our purposes that with the leading-log potential
alone, we can capture the location of the $\Phi_{00}$ VEV~\msvac\ up to small corrections.

In the regime $\epsilon\ll {\alpha_h\over 4\pi}$, the minimum of the potential is indeed located at
$\langle\Phi_{00}\rangle\gg \mu$, which ensures that the approximation used to obtain the minimum is
self-consistent.\foot{Note that this is a stronger assumption than~\reqmi. In fact,~\reqmi\ is sufficient for
the existence of a minimum away from the origin -- since the potential turns up at large fields, as long as the
origin is destabilized such a minimum must exist. In this paper we will only consider the more restricted case
$\epsilon\ll{\alpha_h\over 4\pi}$ where the effective theory is valid. } Another requirement comes from
demanding $\Phi_{00}\ll\Lambda$, so that the effective description in terms of the magnetic theory can be
trusted. According to \msvac, this implies \eqn\reqmii{
 \epsilon \gg {\alpha_{h}\over4\pi} {\mu\over\Lambda} ~.
  }
(In fact, we will see below that an even stronger bound is required for calculability.) Since these requirements
\reqmi, \reqmii\ can obviously be satisfied for the proper choice of parameters, we conclude that a viable
minimum of the potential can be obtained by balancing the two-loop runaway potential against a tree-level mass
term for $\Phi_{00}$.

\subsec{Properties of the metastable vacuum}

Now let us describe the properties of the metastable vacuum in more detail. The deformation~\quarticdef\
explicitly breaks some of the global symmetries~\globalsym. The remaining symmetries are: \eqn\symdef{
SU(N)\times \Big[SU(N_{f0})_D\times SU(N_{f1})\times U(1)_B\times U(1)_1\times U(1)_R\Big]~. } Note that the
choice of $U(1)$ charges in~\globalsym\ makes the symmetry breaking here especially simple.

In the metastable vacuum, the symmetries are broken spontaneously by $\langle \Phi_{00}\rangle\ne 0$ and
$\varphi_1$, $\tilde\varphi_1\ne 0$ to \eqn\sbmsvac{\eqalign{
 & SU(N)\times\Big[ SU(N_{f0})_D\times SU(N_{f1})\times U(1)_B\times U(1)_1\times U(1)_R \Big]\cr
 &\qquad\qquad\qquad \to \Big[SU(N)_D\times SU(N_{f0})_D\times SU(N_{f1}-N)\times  U(1)_B'\times U(1)_1\Big]~.
 }}
Thus we see the crucial difference with~\IntriligatorDD: the $U(1)_R$ symmetry is completely broken in the
metastable vacuum! This can have nice applications to phenomenology, as we will see in the next section.

There is a rich spectrum of fluctuations around the metastable vacuum. We will now describe these, roughly in
order from heaviest to lightest.

\lfm{1.} First of all, the quark superfields $\varphi_{0}$, $\tilde\varphi_0$ are very heavy with $\CO(h
\langle\Phi_{00}\rangle)$ masses. At the tree-level, the squarks have small SUSY splittings $\delta m^2 \sim
h^2\epsilon\mu \langle\Phi_{00}\rangle$.

\lfm{2.} From the description in terms of the effective theory~\efftheory,~\defvace, we see that the spectrum
below the scale $h\langle\Phi_{00}\rangle$ basically factorizes into an ISS-like spectrum for $\Phi_{11}$,
$\varphi_1$, $\tilde\varphi_1$ plus the spectrum of the additional modes $\Phi_{01}$, $\Phi_{10}$, $\Phi_{00}$.
The fields in the ISS sector have masses $\sim h\mu$ (except for the pseudo-moduli, which have one-loop
suppressed masses-squared, and the Goldstone bosons). Some of the fields have large $\delta m^2\sim h^2\mu^2$
splittings. Since these fields have been given a detailed description in~\IntriligatorDD, we will move on to
describing the new fields.

\lfm{3.} Let us split up $\Phi_{01}$ and $\Phi_{10}$ into two submatrices each: \eqn\Phisplit{ \Phi_{01} =
\pmatrix{ \tilde A & \tilde Y}~,\qquad \Phi_{10}=\pmatrix{ A \cr Y}~, } where $A$, $\tilde A^T$ are $N\times
N_{f0}$ and $Y$, $\tilde Y^T$ are $(N_{f1}-N)\times N_{f0}$. Then from the third term in the effective
superpotential \efftheory, we see that $A$, $\tilde A$ acquire masses $\sim h\mu^2/\langle\Phi_{00}\rangle$ from
the VEV of $\varphi_1$. They also have SUSY splittings $\delta m^2\sim {h\mu^2\over
\langle\Phi_{00}\rangle^2}F_{\Phi_{00}}\sim {h^2\epsilon\mu^3\over \langle\Phi_{00}\rangle}$. In order for these
to not make the scalar components of $A$, $\tilde A$ tachyonic, $\langle\Phi_{00}\rangle$ cannot be too large:
$\langle\Phi_{00}\rangle\ll \mu/\epsilon$. Fortunately, this is guaranteed by the formula for the metastable
vacuum~\msvac!

\lfm{4.} Meanwhile $Y$, $\tilde Y$ are massless at tree-level. Since they have a suppressed coupling $\sim
h\mu/\langle\Phi_{00}\rangle$ to the SUSY-split ``messenger" fields in $\varphi_1$, $\tilde\varphi_1$, their
one-loop mass-squared will be similarly suppressed, $m^2\sim ({h\mu\over \langle\Phi_{00}\rangle})^2 (h\mu)^2$.
An explicit calculation shows this mass-squared is positive. Regarding fermions in this
sector, the mass term $\psi_Y\psi_{\tilde Y}$ is allowed by the unbroken symmetries and it is indeed generated
at one-loop. As before, to generate this operator we have to go through propagators which are suppressed by
$1/\langle\Phi_{00}\rangle$.

\lfm{5.} Finally, $\Phi_{00}$ itself obviously has a mass $\sim h\epsilon\mu$, except for ${\rm Im}(\Tr\,\Phi_{00})$ which is the massless Goldstone boson of $U(1)_R$
breaking.\foot{Actually, the effective scalar theory~\defvace\ appears to possess the full $SU(N_{f0})^2$
symmetry even though the superpotential~\quarticdef\ breaks it down to $SU(N_{f0})_D$. So when
$\langle\Phi_{00}\rangle\ne 0$ it would seem there should be many more tree-level massless modes than just the
Goldstone boson of $U(1)_R$ breaking. In fact, these accidental massless modes are all lifted by considering the
most general $SU(N_{f0})_D$-symmetric deformation, instead of just~\quarticdef.}

\bigskip

Finally, let us comment on the potential effects of other quartic operators besides the ones considered in
\quarticdefii. From the discussion of the spectrum, it becomes clear that adding operators of the type
$\epsilon\mu\Phi_{11}\Phi_{10}$, $\epsilon\mu\Phi_{11}\Phi_{01}$ to the superpotential will change our vacuum
significantly. The reason is that the $F$-term of $\Phi_{11}$ gives rise to a tadpole for the light $Y$ modes,
and shifts them far away from the origin. These operators might not change the qualitative conclusions, but for
our analysis to strictly apply they have to be absent for some reason, e.g.\ by a $SU(N_{f0})_D\times
SU(N_{f1})$ global symmetry or discrete symmetries like those we consider in section 4. By the same reasoning,
operators of the form $\epsilon\mu\Phi_{11}\Phi_{00}$ are also dangerous, so they  should also be forbidden by
symmetries. In fact, these operators are automatically forbidden by any symmetry which allows a linear term in
$\Phi_{11}$, while simultaneously forbidding $\Phi_{00}$ and allowing $\Phi_{00}^2$.

\subsec{Calculability, nonperturbative effects and lifetime}

Finally, let us briefly address various issues related
to the consistency of our analysis of the metastable vacuum: calculability, irrelevancy of the nonperturbative
effects, and the lifetime.

First, because the tree-level mass of $\Phi_{00}$ is so much smaller than $\mu$, we need to be careful that
uncalculable corrections to the K\"ahler potential do not overwhelm the tree-level superpotential for
$\Phi_{00}$. In particular, terms like \eqn\kahlerbad{ K_{eff} \supset
{c\over\Lambda^2}\Phi_{11}^\dagger\Phi_{11}\Phi_{00}^\dagger\Phi_{00} } in the K\"ahler potential, with $c$ an
uncalculable $\CO(1)$ number, contribute to the mass of $\Phi_{00}$ via $F_{\Phi_{11}}\sim h\mu^2$:
\eqn\veffbad{ V_{eff}\supset {c\over\Lambda^2}(h\mu^2)^2\Phi_{00}^\dagger\Phi_{00}~. } In order for this to be
negligible compared with the tree-level mass $(h\epsilon\mu)^2$, we clearly need \eqn\sbvaccalc{
 \epsilon \gg {\mu\over\Lambda}~.
 }
Note that this is a {\it stronger} requirement than~\reqmii\ which we deduced above by demanding that
$\langle\Phi_{00}\rangle\ll\Lambda$. We will get back to this constraint in more detail in section 4.

Next, the issue of whether nonperturbative effects can be neglected around the metastable vacuum boils down to
the IR-freedom requirement~\Nfreq, just as in~\IntriligatorDD. In more detail, when $\Phi_{00}$ and $\Phi_{11}$
have generic VEVs, then by the standard scale matching argument the dynamical superpotential is
\eqn\Weff{\eqalign{ W_{dyn}& =  N\left({h^{N_f}\det\,\Phi_{00}\det\,\Phi_{11}\over
\Lambda^{N_{f}-3N}}\right)^{1/N}~. }} The calculability requirement~\Nfreq\ also guarantees that the
non-perturbative superpotential is an irrelevant (dim$>3$) operator around $\Phi_{11}=0$ even for $\Phi_{00}\ne
0$. Thus it can be safely ignored when computing loops around the non-SUSY vacuum.

Finally, let us address the issue of the lifetime. For this we need to know where the SUSY vacua are. At this
point we will take $h=1$ to simplify the expressions, because the value of $h$ does not affect our analysis of
the lifetime. Taking the ansatz $\Phi_{00}\propto \unit_{N_{f0}}$ and $\Phi_{11}\propto \unit_{N_{f1}}$ (and all
other fields zero), it is straightforward to set the $F$-terms of the magnetic superpotential to zero (including
\Weff) and solve. Introducing the additional small parameter \eqn\epsdelt{ \tilde\epsilon ={\mu\over
\epsilon\Lambda}~, } this yields (ignoring phases) \eqn\xysolii{\eqalign{
 \langle\Phi_{00}\rangle_{SUSY} &=\tilde\epsilon^{-{3N_c-2N_f\over 2N_c-N_{f0}}}\epsilon^{ -{3N_c-2N_f+N_c-N_{f0}\over 2N_c-N_{f0}}}\mu=  \tilde\epsilon^{N_{f1}+N\over 2N_c-N_{f0}}\epsilon^{2N\over 2N_c-N_{f0}}\Lambda~,\cr
 \langle \Phi_{11}\rangle_{SUSY}&= \tilde\epsilon^{-{2(3N_c-2N_f)\over 2N_c-N_{f0}}}\epsilon^{-{2(3N_c-2N_f)-N_{f0}\over
 2N_c-N_{f0}}}\mu~.
 }}
So we always have \eqn\Phiineq{
 \mu\ll \langle\Phi_{00}\rangle_{SUSY}\ll \Lambda~,\qquad \langle\Phi_{11}\rangle_{SUSY}\gg\mu~.
 }

Since $\langle\Phi_{11}\rangle_{non-SUSY}=0$,~\Phiineq\ implies that the SUSY vacuum is always separated by an
amount $\Delta\Phi_{11}\gg \mu$ from the non-SUSY vacuum. This is enough to ensure a parametrically long-lived
metastable state, provided that the barrier also scales with $\Delta\Phi_{11}$. The following argument shows
that this must be the case. As in~\IntriligatorDD, the most efficient path between the non-SUSY and SUSY vacua
is to first climb up to a local extremum with $q=\tilde q=0$ and $V= N_{f1}\mu^4$. Then the potential is
schematically (now setting $\Lambda=1$) \eqn\potPhi{ V \sim  N_{f1}\left|
\Phi_{00}^{N_{f0}/N}\Phi_{11}^{N_{f1}/N-1} -\mu^2 \right|^2 + N_{f0}\left|
\Phi_{00}^{N_{f0}/N-1}\Phi_{11}^{N_{f1}/N} -\epsilon\mu \Phi_{00} \right|^2~. } Now the question is how far does
one have to go in field space before the potential decreases to $V_{non-SUSY}= (N_{f1}-N) \mu^4$? Since the
vacuum energy comes primarily from the first term in \potPhi, at the very least we have to decrease it by making
\eqn\lifetimePhi{ \Phi_{00}^{N_{f0}/N}\Phi_{11}^{N_{f1}/N-1}\sim \mu^2~. } Consider now all values of
$\Phi_{00}$ and $\Phi_{11}$ which satisfy \lifetimePhi. We claim that of all such values, either $\Phi_{00}\gg
\mu$ or $\Phi_{11}\gg \mu$. Indeed, if $\Phi_{11}\lesssim \mu$, then $\Phi_{00}\gtrsim \mu^{(3N-N_{f1})/N_{f0}}$
and one can check that this satisfies $\Phi_{00}\gg \langle\Phi_{00}\rangle_{non-SUSY}$, $\langle
\Phi_{00}\rangle_{SUSY} \gg \mu$. Therefore the width of the potential barrier is guaranteed to be
parametrically larger than $\mu$, either through $\Delta\Phi_{11}\gg\mu$ or through $\Delta\Phi_{00}\gg\mu$.
This in turn ensures a parametrically long lifetime for the metastable vacuum.

\subsec{Summary of the requirements on the metastable vacuum}

Since the discussion has been technical at times, we would like to conclude this section by collecting all the
different requirements that need to be satisfied in order for the metastable vacuum to exist.

In addition to requiring that the theory is in the free magnetic phase ($N_c<N_f<{3\over2}N_c$), we also must require
\eqn\constrainti{
 N_{f0} < 3N_c-2N_f \qquad \Longleftrightarrow\qquad N_{f1} > 3(N_f-N_c)
 }
in order for the non-perturbative effects to be negligible around the metastable vacuum, and to ensure a
parametrically long lifetime.

The parameter $\epsilon = m_{\Phi_{00}}/h\mu$ must satisfy the constraints:
\eqn\constraintii{
   {\mu\over\Lambda} \ll   \epsilon \lesssim {\alpha_h\over 4\pi}~.
} The lower bound comes from demanding that uncalculable K\"ahler potential corrections can be neglected
relative to the tree-level mass term. It also ensures that $\langle\Phi_{00}\rangle \ll \Lambda$. The upper
bound comes from requiring that the tree-level mass term does not overwhelm the two-loop potential and stabilize
the metastable vacuum at the origin. We will be assuming a stronger bound $\epsilon\ll{\alpha_h\over 4\pi}$
throughout the paper so that the leading-log approximation to the effective potential~\defvacesub\ can be used.


\newsec{A Simple Model Building Application}

\subsec{The model}

While the main emphasis of our paper is to present a new mechanism of metastable supersymmetry-plus-R-symmetry
breaking, in this section we would also like to briefly touch upon how this mechanism can be applied to building
viable and natural models of gauge mediation. (A more detailed study of the model building applications, in
particular to models of direct gauge mediation, will be reserved for a future publication.) To that end, we will
present here the simplest application of our mechanism to model building: using our SUSY-breaking sector in a
model of  ``ordinary" or ``minimal" gauge mediation~\refs{\DineAG,\DineVC}. Although this may seem like a
trivial application, we will see that various consistency conditions and phenomenological requirements force the
model to be surprisingly predictive.

Our SUSY-breaking sector consists of the theory we studied in great detail in the previous sections. For
simplicity, we assume the minimum number of massless electric quarks, $N_{f0}=1$. We will also assume for
simplicity that $h=1$; in general, it is not a free parameter but is instead determined by various
(uncalculable) $\CO(1)$ numbers coming from Seiberg duality~\IntriligatorDD.

The coupling to the messengers is as in ordinary gauge mediation -- we couple the field $\Phi_{00}$ (which has
both a lowest-component and an $F$-component VEV) to a vector-like pair of messenger fields $\Psi,\t\Psi$
transforming in $5,\bar 5$ of $SU(5)$ GUT: \eqn\Wmess{
   W_{mess} = \lambda\Phi_{00}\Psi\tilde\Psi~.
   }
Here $\lambda$ is some dimensionless coupling; since it is actually a non-renormalizable interaction in the UV,
$\lambda$ will be naturally quite small. We are considering only one messenger pair here; the generalization to
arbitrary numbers of messengers is trivial and does not affect the rough sketch of the physics presented here.

The overall scale of the model is set by phenomenological considerations.  We wish to obtain a realistic
spectrum of SUSY partners, so we set the gaugino mass to be around $100 \ {\rm GeV}$. This is achieved by using
the usual formula of ordinary gauge-mediation \eqn\gmass{
 {\alpha_r\over 4\pi}{F_{\Phi_{00}}\over \Phi_{00}}\sim 10^2\ {\rm
GeV}\quad\Longrightarrow\quad  \epsilon \mu \sim 10^5\ {\rm GeV}~.}

Now let us consider the various consistency conditions on the model. In addition to the
constraints~\constrainti, \constraintii\ on the SUSY breaking sector, we also have the following constraint
coming from the requirement of non-tachyonic messengers: \eqn\constraintiii{
 (\lambda\Phi_{00})^2> \lambda F_{\Phi_{00}}~.
}
Substituting $F_{\Phi_{00}}=\epsilon\mu\Phi_{00}$, this becomes
\eqn\constraintiv{
 \lambda\Phi_{00}>  \epsilon \mu~.
} If $\lambda$ were an $\CO(1)$ coupling then this constraint would be trivial to satisfy, since
$\Phi_{00}\gg\mu$. However, since $\lambda$ comes from a non-renormalizable interaction in the UV, we find that
$\lambda\ll1$ in such a way that~\constraintiv\ becomes quite restrictive, similar to \ShadmiMD.

To see this in more detail, let the $\Phi_{00}^2$ operator be suppressed by a mass scale $M_1$ in the electric
theory, and the $\Phi_{00}\Psi\tilde\Psi$ be suppressed by a (possibly different) mass scale $M_2$. Then we have
\eqn\uvpars{
 \lambda \sim {\Lambda\over M_2}~,\qquad \epsilon\sim {\Lambda^{2}\over M_1\mu}~,
 }
and \constraintiv\ is equivalent to (after substituting $\Phi_{00}\sim  {C\mu\over\epsilon} {\alpha_h\over
4\pi}$ with $C$ the combination of color and flavor factors and the log appearing in \msvac) \eqn\uvparsiii{
  {M_2\over C M_1} < {\mu\over\Lambda\epsilon} {\alpha_h\over 4\pi} \ll 1~.
      }
Thus for $C\sim \CO(1-10)$ there must be a sizeable hierarchy between $M_2$ and $M_1$. In practice this turns
out to be at least a factor of $M_2/M_1\sim 10^{-4}$. This fine-tuning in $M_2$ vs.\ $M_1$ is perhaps the least
attractive feature of this model.\foot{It is interesting to note that, if instead of a quartic deformation, the
leading non-renormalizable term is an even higher-dimension operator, $\delta W\sim {\Lambda^k\over M_1^{2k-3}
}\Phi_{00}^k$, (and assuming the messengers are still coupled through~\Wmess) the constraint of non-tachyonic
messengers becomes trivial for $k\geq 3$. For instance, if $k=3$ the messengers are non-tachyonic as long as
$${M_1\over M_2}>{\Lambda^2\over M_1^2}~.$$ This is a rather
trivial inequality which does not lead to any hierarchy between
$M_1$ and $M_2$.}

We can further manipulate the constraints to derive the allowed ranges of the parameters. We start with \gmass\
and \uvpars, which together imply \eqn\epsmu{ \epsilon \mu \sim {\Lambda^2\over M_1}\sim 10^5\,\,{\rm GeV}~. }
Translating~\constraintii\ to the new parameters leads to \eqn\uvparsiv{
    {\mu\over\Lambda}\ll {\Lambda^2\over M_1\mu}\lesssim {1\over (4\pi)^2}~,
    }
where we have taken $h\sim 1$ for the upper inequality. Combining this with~\epsmu\ we obtain
\eqn\uvparsv{\eqalign{
 & \mu\,\,\gtrsim \,\, 10^7\,\,{\rm GeV}~,\cr
& \Lambda\,\, \gg\,\, {\mu^2\over 10^5\,\,{\rm GeV}} \gtrsim 10^9\,\,{\rm GeV}~,\cr &  M_1 \gg {\mu^4\over
10^{15}\,\,{\rm GeV}^3} \gtrsim 10^{13}\,\,{\rm GeV}~. }}

In fact, if we do not want $M_1$ to exceed the Planck scale, then according to~\uvparsv, $\mu$ is actually
bounded above by $\sim 10^8$ GeV. So we see the parameters can only take values in the ranges \eqn\parranges{
 \mu \sim 10^7{\rm -}10^8\,\,{\rm GeV}~,\qquad  \Lambda \sim 10^9{\rm -}10^{12}\,\,{\rm GeV}~,\qquad M_1\sim 10^{13}{\rm -}10^{19}\,\,{\rm GeV}~.
 }
Apparently, the parameters of the model are quite constrained by all the consistency conditions, and the result
is a relatively predictive version of ordinary gauge mediation!

Let us conclude this subsection by just mentioning one particularly intriguing choice of parameters that
satisfies all the constraints (and in fact minimizes the hierarchy between $M_2$ and $M_1$): \eqn\onechoice{
 \mu\sim 10^8 \,\,{\rm GeV}~,\qquad  \Lambda \sim 10^{12}\,\,{\rm GeV},
 \qquad M_1\sim \pl \sim 10^{19}\,\,{\rm GeV}~.
 }
This satisfies all the requirements as long as $M_2\lesssim 10^{15}$ GeV. Interestingly, this is tantalizingly
close to the GUT scale, so perhaps one could imagine a UV completion where GUT-scale fields connect $\Phi_{00}$
to the messengers, but only Planck-suppressed interactions generate $\Phi_{00}^2$.

\subsec{Phenomenological features of the model}

Now let us briefly mention some phenomenological aspects of our ordinary gauge mediation model. For concreteness
we will focus on the point~\onechoice\ with $M_2\sim 10^{15}$ GeV. However, we would like to emphasize that the
statements we will make in this subsection are for the most part {\it predictions} of our model, not simply
consequences of the particular choice of parameters.

In our model, the parameters of ordinary gauge mediation are
\eqn\ogmpars{\eqalign{
  & M_{mess} \sim \lambda\Phi_{00} \sim 10^5{\rm-}10^6\,\,{\rm GeV}~,\cr
  & F\sim \lambda F_{\Phi_{00}}
  \sim 10^{10}{\rm -}10^{11}\,\,{\rm GeV}^2~,\cr
  & m_{3/2}\sim {F_{\Phi_{11}}\over M_{\rm pl}} \sim 1{\rm-}10\,\, {\rm MeV}~.
  }}
Here we have used $\lambda\sim 10^{-3}$ and $\epsilon\sim
10^{-3}$. Note that $M_{mess}$ and $F$ come from the same source,
the R-symmetry breaking scale $\Phi_{00}$, and therefore there are
no extra CP phases in the gaugino masses even if doublet/triplet
splitting of the messengers is present. Moreover, having one and
the same source for $M_{mess}$ and $F$, as we do, guarantees
messenger-parity, which is necessary for a consistent spectrum of
soft terms.

Since R-symmetry is spontaneously broken, the R-axion is massless in gauge theory, up to small non-perturbative
contributions. On top of the field theory effects, there is also a supergravity contribution to the mass $m_a$.
Following~\BaggerHH\ we find that this contribution is \eqn\maxion{ m_a\sim \sqrt{{\epsilon\mu^3\over M_{\rm
pl}}}\sim 10 \,\,{\rm GeV}~. } This is already well above the astrophysical bound, $m_a\gtrsim 10 \,\, {\rm
MeV}$.

Note that the mass of the gravitino in~\ogmpars\ is always determined by the largest $F$-term in the
SUSY-breaking sector. In this model, this is not $F_{\Phi_{00}}$ but is instead $F_{\Phi_{11}}\sim\mu^2\sim
10^{16}$ GeV$^2$. So what we have is a model with light messengers but a heavy ($\sim 1\ {\rm MeV}$) gravitino.
This combination of features is generally rare in models of gauge-mediation; here it is achieved because the
hidden sector has a subsector with much larger SUSY-breaking than the SUSY-breaking felt by the
messengers.\foot{The gravitino mass also sets the scale of all the gravity mediated contributions to soft
terms. These contributions are not degenerate and, in general, violate flavor in the maximal possible way. For
the gravitino masses considered here, there is no contradiction with precision measurements and these
contributions are allowed to be arbitrary. For a recent work analyzing flavor aspects of combined gauge and
gravity mediation see~\FengKE.}

A heavy gravitino can have observable consequences at colliders. In general, a gravitino mass of~\ogmpars\
implies that the NLSP is sufficiently long lived that it will escape the detector. (For a good review of this
subject and more references, we refer the reader to~\GiudiceBP.) Thus the standard $\gamma\gamma+$MET signal of
gauge mediation will not be observed. On the other hand, if the NLSP is a very long-lived slepton (as would be
the case, for instance, if the messenger scale is very low or there are $N_m\sim 5$ messengers) then the
collider signatures of the model could be quite spectacular.

\subsec{Naturalizing the UV model}

To build a natural model (starting from some high scale) with the dynamics described above we clearly need to
forbid all the super-renormalizable terms in the UV Lagrangian, while still allowing for the quartic
interactions between the massless electric quarks. In addition, as explained in section~3, some quadratic
operators must be forbidden for our analysis to be reliable. We would also like to forbid the most general
couplings between the electric quarks and the messengers while allowing for~\Wmess; otherwise the messenger SUSY
masses and $F$-terms will no longer necessarily come from the same source. Finally, we would like to generate
the mass scale $\mu$ dynamically. As we will see in this section, all four of these conditions can be met by
``retrofitting the model"~\DineGM\ with an auxiliary gauge group and imposing its associated non-anomalous
discrete R-symmetry. This approach was implemented recently in a similar MGM-type context by~\AharonyMY.

The main idea of retrofitting
is as follows \DineGM.  Consider a pure super Yang-Mills (SYM) theory
$SU(\tilde N)$ with the strong scale being $\tilde \Lambda$. At energies below $\tilde \Lambda$ there is a
gaugino condensate which can be used to source masses for some of the electric quarks. Furthermore, the SYM theory has a non-anomalous ${\Bbb Z}_{2\tilde N}$ R-symmetry, and this can be promoted to a
gauged discrete symmetry of the entire theory.

Now let us turn to a concrete realization of the UV theory along these lines. Consider the following gauge
symmetry: \eqn\gauge{SU(3)\times SU(N_c)\times G_{SM} \times {\Bbb Z}_6~.} The $SU(3)$ group is a pure SYM
theory, $G_{SM}$ is the SM (or GUT) gauge group and $SU(N_c)$ is the gauge group of SQCD. Let $Q_1$ denote the
$N_{f1}$ massive electric quarks of SQCD and $Q_0$ be the only electric quark which is massless in the UV.
${\Bbb Z}_6$ is generated by $g$ such that $g^6=1$. We assign the following charges under ${\Bbb Z}_6$:
\eqn\charges{ \int d^2\theta :\,\,\, g^{-2}~;\qquad W_\alpha W^\alpha,\,\,Q_0,\,\,\tilde
Q_0,\,\,\Psi,\,\,\tilde\Psi:\,\,\, g^2~;\qquad Q_1,\,\,\tilde Q_1:\,\,\, g^3~. } Here $W_\alpha$ is the chiral
field strength of the $SU(3)$ SYM. Note that these charge assignments forbid all the renormalizable couplings.
Now let us write down the first few terms in the superpotential of this theory: \eqn\uvsup{ W = {1 \over M_1}
(Q_0 \tilde Q_0)^2 +{1\over M_2} Q_0 \tilde Q_0 \Psi \tilde \Psi  +{1\over M_3^2 } {\rm Tr } W_\alpha W^\alpha
Q_1 \tilde Q_1+\dots~. } Here $\dots$ stand for operators of dimension $6$ and higher, as well as for quartic
messenger self-interactions, both of which are irrelevant for our discussion. We see that the dangerous terms
$Q_1\tilde Q_1 \Psi\tilde\Psi$ are forbidden by the discrete R-symmetry. We also see that all but the
$\Phi_{00}^2$-type quartic interactions are forbidden.\foot{We wish to emphasize again that only {\it some} of the additional quadratic operators cause significant shifts of the vacuum we analyzed (and even in this case it could very well be that our qualitative conclusions remain intact), so the fact {\it all} of them are forbidden by the symmetries~\gauge\ is an added (but by no means necessary) benefit.  }

To estimate $\tilde \Lambda$ we take for simplicity $M_3\sim \pl$. For $\mu^2 \sim 10^{16}\ {\rm
GeV}^2$~\onechoice, we find $\tilde \Lambda \sim 10^{14}\ {\rm GeV}$. Hence $\tilde \Lambda\gg\Lambda$, which is
necessary for the consistency of our analysis. Gaugino condensation breaks ${\Bbb Z}_6$ to ${\Bbb Z}_2$, which
is an unbroken symmetry of the theory. Indeed, $\Phi_{00} \sim Q_0 \tilde Q_0$, which obtains a VEV, is R-even. Note that this ${\Bbb Z}_2$ R-symmetry is naturally extended to the R-parity of the MSSM.

Lastly, let us briefly discuss the issue of the discrete R-symmetry anomalies. The classification and analysis
of the various possible anomalies has been discussed, for instance, in~\refs{\KraussZC,\IbanezHV,\IbanezJI}. In
particular,~\IbanezJI\ argued that the cubic anomaly does not impose any constraints on the low energy effective
action. So, we remain with the constraints of gravitational anomaly and mixed gauge-discrete anomalies.  The
mixed anomaly with the strong $SU(3)$ is automatically satisfied and the mixed anomaly with $SU(N_c)$ gives
\eqn\mixed{2N_{f1}+N_{f0} =3r~,} where $r$ is some integer.

On the other hand, since we have not specified the R-charges of the SSM fields and have not even chosen the
matter content,\foot{In particular, to solve the $\mu/B\mu$ problem one usually introduces many new fields
(e.g.~\GiudiceCA ) and various special mutual interactions. The particular choice of the mechanism to solve
$\mu/B\mu$ may affect the charge assignments of the potentially observable particles.} we are unable to
calculate the gravitational anomaly and the mixed anomaly with the SSM gauge group. The specific charge
assignment here should be merely considered as an example of how to implement the above mentioned idea of
naturalization, and we leave a more detailed analysis to the future.

\bigskip

\bigskip

\bigskip

\noindent {\bf Acknowledgments:}

We would like to thank O.~Aharony, Y.~Hochberg, Y.~Nir and Y.~Shadmi for stimulating discussions. We are
grateful to the organizers of ``String Theory - From Theory to Experiment" at the Institute for Advanced Studies
in the Hebrew University, where this work was initiated. The work of A.~G. is supported in part by the BSF --
American-Israel Bi-National Science Foundation, by a center of excellence supported by the Israel Science
Foundation (grant number 1468/06), EU grant MRTN-CT-2004-512194, DIP grant H.52, and the Einstein Center at the
Hebrew University. The work of A.~K. is supported in part by the Israel-U.S. Binational Science Foundation (BSF)
grant No. 2006071 and by Israel Science Foundation (ISF) under grant 1155/07. The work of Z.~K. is supported in
part by the Israel-U.S. Binational Science Foundation, by a center of excellence supported by the Israel Science
Foundation (grant number 1468/06), by a grant (DIP H52) of the German Israel Project Cooperation, by the
European network MRTN-CT-2004-512194, and by a grant from G.I.F., the German-Israeli Foundation for Scientific
Research and Development. The work of D.~S. was supported in part by NSF grant PHY-0503584. Any opinions,
findings, and conclusions or recommendations expressed in this material are those of the author(s) and do not
necessarily reflect the views of the National Science Foundation.

\appendix{A}{Calculating the Runaway Potential at Large Fields}

For the sake of completeness, we will describe in this appendix the RGE evolution of the wavefunction of
$\Phi_{11}$ for large $\Phi_{00}$. A more general derivation of the leading-log potential for pseudo-moduli at
large fields can be found in~\ISSiii.

We will assume that $\langle\Phi_{00}\rangle=z\unit_{N_{f0}}$ for simplicity; the generalization to arbitrary
non-degenerate $\Phi_{00}$ will be obvious at the end. The wavefunction factors in the K\"ahler potential are
governed by the RGEs: \eqn\wvfnrges{ \log \CZ_{\Phi_{11}}= 2\int_{Q}^{\Lambda}\,{dQ'\over
Q'}\gamma_{\Phi_{11}}(Q';z)~, } where $\gamma_{\Phi_{11}}$ is its anomalous dimension, and $Q$ is the RG scale.
In this model, the relevant one-loop anomalous dimensions above the scale $hz$ are \eqn\gammaXspecabove{
\gamma_{\Phi_{11}} = {Nh^2\over 16\pi^2\CZ_{\Phi_{11}}\CZ_{\varphi_1}^2}~,\qquad
\gamma_{\varphi_1}=\gamma_{\tilde\varphi_1}={N_{f1}h^2\over 16\pi^2\CZ_{\Phi_{11}}\CZ_{\varphi_1}^2}
+{N_{f0}h^2\over 16\pi^2\CZ_{\Phi_{10}}\CZ_{\varphi_0}\CZ_{\varphi_1}}~, } and below the scale $hz$ they are
\eqn\gammaXspecbelow{ \gamma_{\Phi_{11}} =  {Nh^2\over 16\pi^2\CZ_{\Phi_{11}}\CZ_{\varphi_1}^2}~,\qquad
\gamma_{\varphi_1}=\gamma_{\tilde\varphi_1}={N_{f1}h^2\over 16\pi^2\CZ_{\Phi_{11}}\CZ_{\varphi_1}^2}~. } In
particular, $\gamma_{\Phi_{11}}$ is continuous at the scale $h z$. Therefore we have \eqn\wvfnPhidep{
{\partial\log \CZ_{\Phi_{11}}\over \partial\log hz} = 2\int_{Q}^{\Lambda}{dQ'\over Q'}
{\partial\gamma_{\Phi_{11}}(Q';z)\over\partial\log hz}= -2\int_Q^\Lambda {dQ'\over
Q'}\,\gamma_{\Phi_{11}}{\partial(\log\CZ_{\Phi_{11}}+2\log \CZ_{\varphi_1})\over\partial\log hz}~. } At
leading-log order, only the second term in \wvfnPhidep\ matters, and only through the discontinuity in
$\gamma_{\varphi}$: \eqn\discvarphi{ \log\CZ_{\varphi_1} =
 \int_{hz}^{\Lambda}{dQ'\over Q'}\left( {N_{f1}h^2\over 8\pi^2\CZ_{\Phi_{11}}\CZ_{\varphi_1}^2} +{N_{f0}h^2\over 8\pi^2\CZ_{\Phi_{10}}\CZ_{\varphi_0}\CZ_{\varphi_1}}\right)  +  \int_{Q}^{hz}{dQ'\over Q'}\left( {N_{f1}h^2\over 8\pi^2\CZ_{\Phi_{11}}\CZ_{\varphi_1}^2} \right) \qquad
 (Q<hz)~,
 }
so \eqn\hiisolii{ {\partial \log \CZ_{\varphi_1}\over\partial \log hz} = -{N_{f0} h^2\over
8\pi^2\CZ_{\Phi_{10}}\CZ_{\varphi_0}\CZ_{\varphi_1}} \Theta(hz-Q)+\dots~, } where $\dots$ do not contribute in
the leading log approximation. Substituting this into \wvfnPhidep\ and integrating twice, we conclude that
 \eqn\wvfnPhidepiii{\eqalign{ \log \CZ_{\Phi_{11}}
&={Nh^2\over 8\pi^2}{N_{f0} h^2\over 8\pi^2}\left(\log{hz\over Q}\right)^2 +\dots\cr }~,} where we have dropped
the wavefunction factors since they will only contribute at higher loop order.

Now, it is trivial to generalize to arbitrary $\Phi_{00}$. For instance, if $\Phi_{00}$ is diagonal but has
different eigenvalues, the same arguments as above lead to \eqn\wvfnPhidepiv{ \log\CZ_{\Phi_{11}} ={Nh^2\over
8\pi^2}{ h^2\over 8\pi^2}\sum_{i=1}^{N_{f0}}\left(\log{hz_i\over Q}\right)^2 +\dots~. } Since $\Phi_{00}$ can
always be diagonalized by an $SU(N_{f0})_L\times SU(N_{f0})_R$ transformation, the full generalization of
\wvfnPhidepiii, \wvfnPhidepiv\ must be \eqn\wvfnPhidepv{ \log\CZ_{\Phi_{11}} ={Nh^2\over 16\pi^2}{ h^2\over
16\pi^2}\Tr\,\left(\log{h^2\Phi_{00}^\dagger\Phi_{00}\over Q^2}\right)^2 + \dots~, } which is the result quoted
in the text upon substituting $Q=h\mu$.

 \listrefs
\end